\documentclass[%
 reprint,
superscriptaddress,
 amsmath,amssymb,
 aps,
 prb,
showkeys
]{revtex4-2}

\usepackage{graphicx}
\usepackage{dcolumn}
\usepackage{bm}
\usepackage[hidelinks]{hyperref}
\usepackage{rotating}
\usepackage{booktabs}
\usepackage{threeparttable}
\usepackage{subcaption}

\usepackage{soul}
\usepackage{color}
\usepackage[usenames,dvipsnames]{xcolor}
\newcommand{\hlc}[2][yellow]{ {\sethlcolor{white} \hl{#2}} }

\makeatletter
\renewcommand{\fnum@figure}{Figure \thefigure}
\makeatother

\begin{document}

\preprint{APS/123-QED}

\title{Machine learning spectral indicators of topology}

\author{Nina Andrejevic}
\email[These authors contributed equally to this work. \\Corresponding author: ]{nandrejevic@anl.gov}
\affiliation{Center for Nanoscale Materials, Argonne National Laboratory, Lemont, IL 60439, USA}
\affiliation{Quantum Measurement Group, Massachusetts Institute of Technology, Cambridge, MA 02139, USA}
\affiliation{Department of Materials Science and Engineering, Massachusetts Institute of Technology, Cambridge, MA 02139, USA}
\author{Jovana Andrejevic}
\email[These authors contributed equally to this work. \\Corresponding author: ]{nandrejevic@anl.gov}
\affiliation{Department of Physics, University of Pennsylvania, Philadelphia, PA 19104, USA}
\affiliation{John A.~Paulson School of Engineering and Applied Sciences, Harvard University, Cambridge, MA 02138, USA}
\author{B. Andrei Bernevig}
\affiliation{Department of Physics, Princeton University, Princeton, NJ 08544, USA}
\affiliation{Donostia International Physics Center, P. Manuel de Lardizabal 4, Donostia-San Sebastian, 20018, Spain}
\affiliation{IKERBASQUE, Basque Foundation for Science, Plaza Euskadi 5, Bilbao, 48009, Spain}
\author{Nicolas Regnault}
\affiliation{Department of Physics, Princeton University, Princeton, NJ 08544, USA}
\author{Fei Han}
\affiliation{Quantum Measurement Group, Massachusetts Institute of Technology, Cambridge, MA 02139, USA}
\affiliation{Department of Nuclear Science and Engineering, Massachusetts Institute of Technology, Cambridge, MA 02139, USA}
\author{Gilberto Fabbris}
\affiliation{Advanced Photon Source, Argonne National Laboratory, Lemont, IL 60439, USA}
\author{Thanh Nguyen}
\affiliation{Quantum Measurement Group, Massachusetts Institute of Technology, Cambridge, MA 02139, USA}
\affiliation{Department of Nuclear Science and Engineering, Massachusetts Institute of Technology, Cambridge, MA 02139, USA}
\author{Nathan C. Drucker}
\affiliation{Quantum Measurement Group, Massachusetts Institute of Technology, Cambridge, MA 02139, USA}
\affiliation{John A.~Paulson School of Engineering and Applied Sciences, Harvard University, Cambridge, MA 02138, USA}
\author{Chris H. Rycroft}
\email[Corresponding author: ]{chr@math.wisc.edu}
\affiliation{Department of Mathematics, University of Wisconsin-Madison, Madison, WI 53706, USA}
\affiliation{John A.~Paulson School of Engineering and Applied Sciences, Harvard University, Cambridge, MA 02138, USA}
\affiliation{Computational Research Division, Lawrence Berkeley Laboratory, Berkeley, CA 94720, USA}
\author{Mingda Li}
\email[Corresponding author: ]{mingda@mit.edu}
\affiliation{Quantum Measurement Group, Massachusetts Institute of Technology, Cambridge, MA 02139, USA}
\affiliation{Department of Nuclear Science and Engineering, Massachusetts Institute of Technology, Cambridge, MA 02139, USA}

\date{\today}
\begin{abstract}
Topological materials discovery has emerged as an important frontier in condensed matter physics. While theoretical classification frameworks have been used to identify thousands of candidate topological materials, experimental determination of materials’ topology often poses significant technical challenges. X-ray absorption spectroscopy (XAS) is a widely-used materials characterization technique sensitive to atoms' local symmetry and chemical bonding, which are intimately linked to band topology by the theory of topological quantum chemistry (TQC). Moreover, as a local structural probe, XAS is known to have high quantitative agreement between experiment and calculation, suggesting that insights from computational spectra can effectively inform experiments. In this work, we leverage computed X-ray absorption near-edge structure (XANES) spectra of more than 10,000 inorganic materials to train a neural network (NN) classifier that predicts topological class directly from XANES signatures, achieving F$_1$ scores of 89\% and 93\% for topological and trivial classes, respectively. \hlc[cyan]{Additionally, we obtain consistent classifications using corresponding experimental and computational XANES spectra for a small number of measured compounds.} Given the simplicity of the XAS setup and its compatibility with multimodal sample environments, the proposed machine learning-augmented XAS topological indicator has the potential to discover broader categories of topological materials, such as non-cleavable compounds and amorphous materials, and may further inform field-driven phenomena \textit{in situ}, such as magnetic field-driven topological phase transitions.
\end{abstract}

\keywords{machine learning, topological materials, X-ray absorption spectroscopy}
\maketitle

\section{Introduction}
Topological materials are characterized by a topologically nontrivial electronic band structure from which they derive their exceptional transport properties~\cite{RevModPhys.82.3045,RevModPhys.83.1057,yan2012topological,RevModPhys.88.021004,yan2017topological,RevModPhys.90.015001}. The prospect of developing these exotic phases into useful applications has garnered widespread efforts to identify and catalogue candidate topological materials, evidenced by the emergence of numerous theoretical frameworks based on connectivity of electronic bands~\cite{bradlyn2017topological,kruthoff2017topological,cano2018building,elcoro2020application,wieder2021topological,bouhon2021topological,cualuguaru2021general}, symmetry-based indicators ~\cite{slager2013space,jadaun2013topological,chiu2016classification,bradlyn2017topological,po2017symmetry,song2018quantitative,song2019topological,po2020symmetry,peng2021topological}, electron-filling constraints~\cite{bradlyn2017topological,chen2018topological,watanabe2018structure}, and spin–orbit spillage \cite{choudhary2019high,choudhary2020computational,choudhary2021high}. These frameworks have facilitated the prediction of over 8,000 topologically non-trivial phases~\cite{vergniory2019complete,zhang2019catalogue,tang2019comprehensive,tang2019topological,tang2019efficient,wang2019two,xu2020high,vergniory2021all}, a vast unexplored territory for experiments. This is strong motivation to develop complementary experimental techniques for high-throughput screening of candidate materials. Current state-of-the-art techniques such as angle-resolved photoemission spectroscopy (ARPES), scanning tunneling microscopy (STM), and quantum transport measurements are commonly used to detect topological signatures, but a few limitations remain: Methods like ARPES directly probe band topology but are surface-sensitive and thereby place strict requirements on sample preparation and the sample environment, limiting the range of experimentally accessible materials~\cite{suga2013photoelectron,lv2019angle}; transport measurements, on the other hand, can be performed on more versatile samples but can be more difficult to interpret. Neither approach yet fully meets the demands of a high-throughput classification program.

Machine learning methods are increasingly being adapted to materials research to accelerate materials discovery~\cite{raccuglia2016machine,liu2017materials,gomez2018automatic,zhang2019thermal,mikulskis2019toward,juan2020accelerating,juan2020accelerating,kusne2020fly,mannodi2021computational} and facilitate inverse design through high-throughput property prediction~\cite{pilania2013accelerating,ward2016general,carrete2014finding}. Several recent studies have proposed data-driven frameworks for predicting band topology from structural and compositional attributes~\cite{claussen2019detection,rodriguez2019identifying,ma2022topogivity} and quantum theoretical or simulated data~\cite{PhysRevLett.118.216401,PhysRevLett.122.210503,scheurer2020unsupervised,zhang2018machine}. At the same time, machine learning methods are being adopted to automate and improve data analysis for a broad range of experimental techniques ~\cite{RevModPhys.91.045002,PhysRevMaterials.3.033604,PhysRevApplied.12.054049,han2019deep,samarakoon2019machine,zhang2019machine,rem2019identifying}. Importantly, machine learning presents a potential opportunity to not only accelerate data analysis, but to derive useful information from complex data in the absence of reliable theoretical models, or to extract new insights beyond traditional models.

In this work, we develop a data-driven classifier of electronic band topology using materials' X-ray absorption spectra. X-ray absorption spectroscopy (XAS) is widely used to characterize the chemical state and local atomic structure of atomic species in a material. This technique is suitable for the study of highly diverse samples and environments, including noncrystalline materials and extreme temperatures and pressures \cite{newville2014fundamentals}. As a bulk probe, XAS also places few constraints on surface quality and sample preparation. The X-ray absorption near-edge structure (XANES), defined within approximately 50 eV of an XAS absorption edge, provides a specie-specific fingerprint of the absorbing atom's local chemical environment, including coordination chemistry, orbital hybridization, and density of available electronic states. However, despite the rich electronic structural information contained in XANES spectra, the lack of a simple analytic description of XANES has compelled largely qualitative treatment of this energy regime, with individual spectral features attributed to properties of the electronic structure through empirical evidence and spectral matching \cite{gaur2015speciation}. As a result, machine learning methods have been introduced to automate the estimation of materials parameters such as coordination environments \cite{PhysRevMaterials.3.033604,torrisi2020random,zheng2020random,kiyohara2018data,guda2021understanding}, oxidation states, \cite{torrisi2020random,guda2021understanding}, and crystal-field splitting \cite{suzuki2019automated} from XANES and other core-level spectroscopies, and even enable direct prediction of XANES spectra from structural and atomic descriptors \cite{carbone2020machine,rankine2020deep,lueder2021machine}. Here, we propose that machine learning models can be used to extract other hidden electronic properties, namely the electronic band topology, from XANES signatures and thereby serve as a potentially useful diagnostic of topological character. The theory of topological quantum chemistry (TQC) has demonstrated the intimate link between a material's band topology and its local chemical bonding~\cite{bradlyn2017topological}, which motivates our inquiry into the unexplored connection between XANES spectra and band topology. In particular, we develop a machine learning-enabled indicator of band topology based on K-edge XANES spectral inputs, which correspond to electronic transitions from the $1s$ core shell states to unoccupied states above the Fermi energy. First, we summarize the data assembly procedure, which consists of labeling the database of computed XANES K-edge spectra~\cite{mathew2018high} according to topological character using the catalogue of high-quality topological materials predicted by TQC~\cite{vergniory2019complete,vergniory2021all}. We then conduct an exploratory analysis of topological indication for the K-edge XANES spectra of different elements based on principal component analysis (PCA) and $k$-means clustering. Finally, we develop a neural network (NN) classifier of topology that synthesizes insights from XANES signatures of all elements in a given compound. Our classifier achieves F$_1$ scores of $89\%$ and $93\%$ for topological and trivial classes, respectively. Materials containing certain elements, including Be, Al, Si, Sc, Ti, Ga, Ag, and Hg, are predicted with F$_1$ scores above $90\%$ in both classes. Our work suggests the potential of machine learning to uncover topological character embedded in complex spectral features, especially when a mechanistic understanding is challenging to acquire.

\begin{figure*}[!t]
  \centering
  \includegraphics[width=16cm]{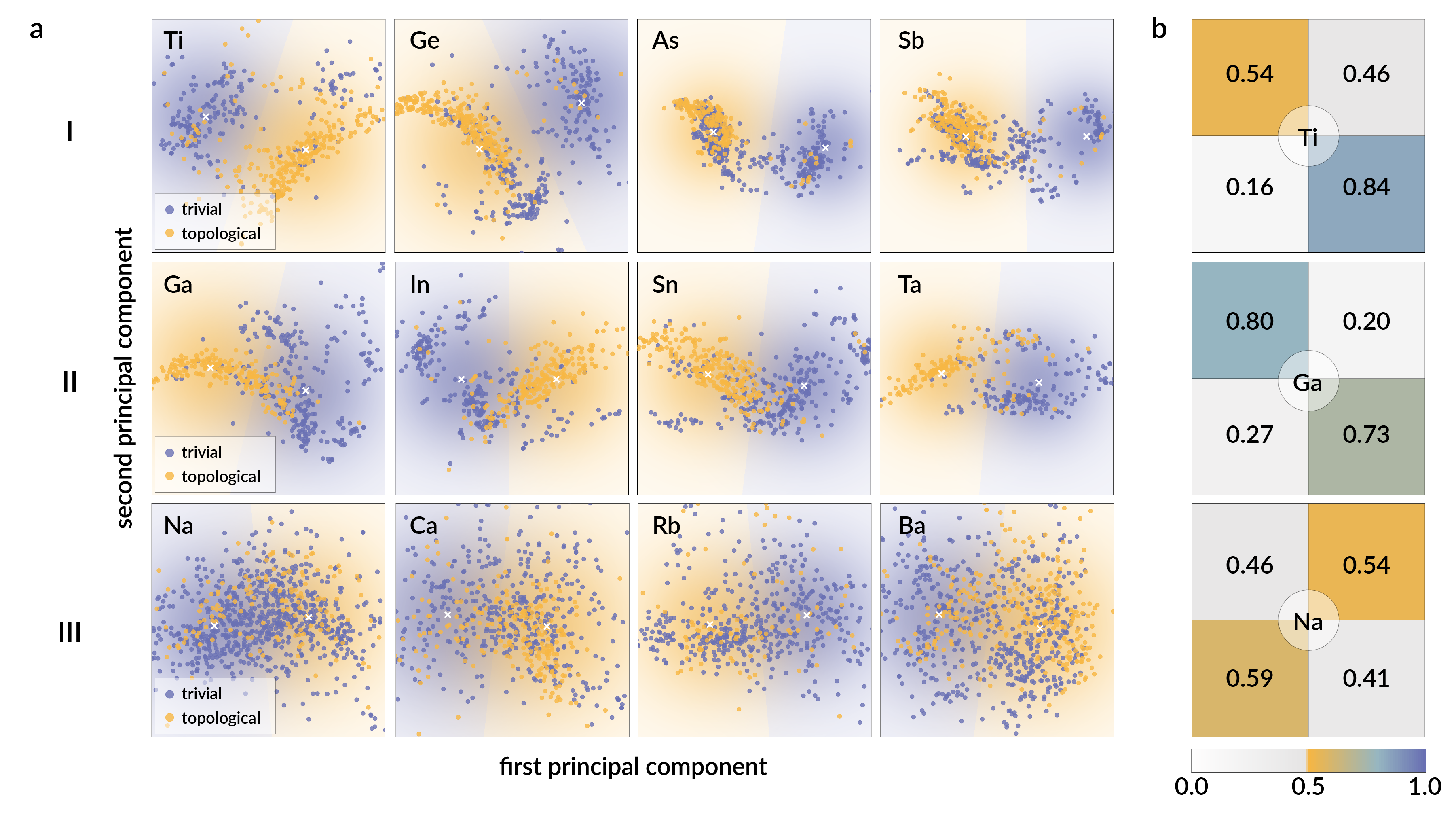}
  \caption{\raggedright\textbf{Exploratory analysis using principal components and \textit{k}-means clustering.} (a) Decision boundary visualizations of classifications by unsupervised \textit{k}-means clustering for selected elements. As detailed in the main text, the \textit{k}-means clustering is performed on the subset of principal components accounting for at least $80\%$ of the explained variance of spectra for a given element. The clusters are visualized along the first ($x$-axis) and second ($y$-axis) principal components in the scatter plots. Scattered points are colored according to their true class: topological (orange) or trivial (blue). The background is shaded according to the cluster-assigned class. The principal components exhibited three typical patterns: (row I) imbalanced classification in favor of topological examples, (row II) relatively balanced classification of topological and trivial examples, and (row III) no apparent clustering by class. (b) Confusion matrices of representative examples in each of rows I, II, and III.}
  \label{fig:figure1}
\end{figure*}

\begin{figure*}[!t]
  \centering
  \includegraphics[width=18cm]{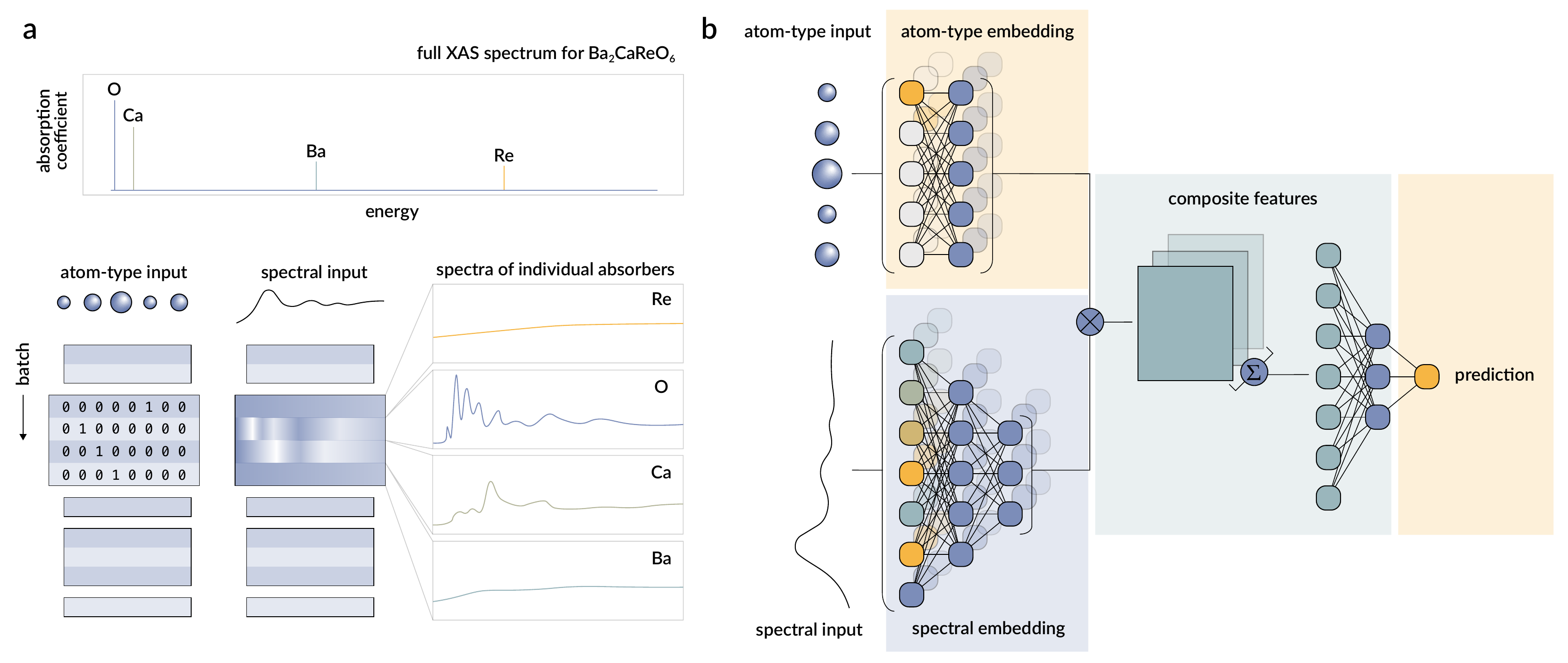}
  \caption{\raggedright\textbf{Data structure and model architecture.} (a) A schematic of the full XANES spectrum for a representative sample in the dataset, showing the signatures from different absorbing elements on an absolute energy scale. For a given material, the inputs to the NN classifier consist of one-hot encoded atom types (left) and XANES spectra (right) for all absorbing atoms. (b) Schematic of the NN architecture predicting the (binary) topological class using spectral and atom-type inputs. Spectral and atom-type inputs are individually embedded by fully-connected layers before performing a direct product between corresponding spectral and atomic channels. These composite features are aggregated for a given material and passed to a final fully-connected block to predict the topological class.}
  \label{fig:figure2}
\end{figure*}

\section{Data preparation and pre-processing}
XAS data were obtained from the published database of computed K-edge XANES spectra~\cite{mathew2018high} and additional examples distributed on the Materials Project~\cite{Jain2013,10.1038/s41524-018-0067-x,Ong2012b,Ong_2015}, which are computed using the FEFF9 program \cite{rehr2010parameter}. The materials from the XANES database were then labeled according to their classification in the database of topological materials~\cite{vergniory2019complete,vergniory2021all}, which is based on the formalism of TQC~\cite{bradlyn2017topological}. The classifications in the TQC database are based on structures from the Inorganic Crystal Structure Database (ICSD)~\cite{bergerhoff1987crystallographic}, and the ICSD identifier was used to associate topological class labels with entries in the XANES database. We note that the crystal structures in the two databases are not strictly identical, and ICSD identifiers are associated with structurally-similar Materials Project entries according to pymatgen's StructureMatcher algorithm \cite{Ong_2015,Ong2012b}. \hlc[yellow]{In rare cases, multiple ICSD identifiers corresponding to different topological classifications were associated with the same set of XANES spectra. Because small discrepancies between the ICSD and Materials Project structures could lead to different topological classification for some materials close to a phase transition, all multiply-labeled examples were removed from the dataset.} The materials data were further refined based on availability of both high-quality topological classification and spectral data, resulting in $13,151$ total materials considered: $4,957$ topological $(\sim38\%)$ and $8,194$ trivial $(\sim62\%)$. Here, high-quality is defined following Ref. \citenum{vergniory2019complete}, which considers only materials with well-determined structures and excludes alloys, magnetic compounds, and certain problematic $f$-electron atoms. Additionally, entries with spectra containing unphysical features such as large negative jumps were discarded. The materials in the final dataset are structurally and chemically diverse, representing 200 of 230 spacegroups and 63 different elements, with primitive unit cells ranging from 1 to 76 atoms and up to 7 unique chemical species. The representation of different elements among topological and trivial examples is shown in \textbf{Figure~\ref{fig:figureS1}a-b} of the Supplemental Material. Data were subdivided into training, validation, and test sets according to a $70/15/15\%$ split. While samples were randomly distributed among the datasets, an assignment process was developed to ensure balanced representation of each absorbing element and topological class within each dataset. Specifically, the fraction of topological insulators (TI), topological semimetals (TSM) and topologically trivial materials represented in compounds containing a certain element was balanced as shown in \textbf{Figure~\ref{fig:figureS1}c}. For each example, the computed K-edge XANES spectra of each absorbing element were interpolated and re-sampled at $200$ evenly-spaced energy values spanning an energy range of 56 eV surrounding the absorption edge. The spectra were standardized separately for different absorbing elements, which consisted of centering the mean of spectral intensities over each energy range, and scaling by the average intensity standard deviations.

\section{Results}
\subsection{Exploratory analysis}
Prior to training the NN classifier, we conducted an exploratory analysis of the assembled XANES spectra to estimate the separability by topological class exhibited by different elements. For all examples containing a given element, we performed a principal component analysis (PCA) on the high-dimensional spectra and subsequently carried out unsupervised \textit{k}-means clustering on a subset of principal components of the training set. The number of retained principal components was selected to retain at least $80\%$ of the explained variance of spectra for a given element. Results of the clustering analysis for a selection of elements are shown in \textbf{Figure~\ref{fig:figure1}}. The decision boundary between the two clusters identified by \textit{k}-means clustering, projected along the first two principal components, lies at the intersection of the blue (trivial) and orange (topological) shaded regions in \textbf{Figure~\ref{fig:figure1}a}. Since \textit{k}-means clustering is not supervised by the true topological class of each example, cluster assignment was performed by solving an optimal matching problem that finds the pairing between clusters and topological classes that minimizes the number of misclassified examples, corrected for class imbalance. The examples from all three datasets (training, validation, and testing) are plotted as scattered points in the low-dimensional space and colored according to their known topological class. Additional visualizations are shown in \textbf{Figure~\ref{fig:figureS2}}. A quick survey of these results reveals a number of elements for which the classification accuracy of topological and trivial examples is imbalanced, and a few for which the classification accuracy is more balanced between the two classes. We correlated these observations with the decision boundary visualizations and noted three distinct patterns in the result of our unsupervised clustering. For some elements, nearly all topological examples were segregated within a single cluster (row I of \textbf{Figure~\ref{fig:figure1}}). This led to a strong score for topological examples but weaker score for trivial ones for elements such as Ti, Ge, As, and Sb. Other elements like Ga, In, Sn, and Ta exhibited more balanced classification accuracies between the two topological classes (row II of \textbf{Figure~\ref{fig:figure1}}). On the other hand, there were a number of unsuccessful examples of alkali and alkaline earth metals for which clustering of the data did not appear coincident with topological class (row III of \textbf{Figure~\ref{fig:figure1}}). Given that the feature transformations performed in our exploratory analysis were element-specific, the potential to discriminate data between the two classes is encouraging. This also suggests a possible advantage of synthesizing information of all constituent atom types in a given compound in order to improve prediction accuracy.

\begin{figure*}[!t]
  \centering
  \includegraphics[width=18cm]{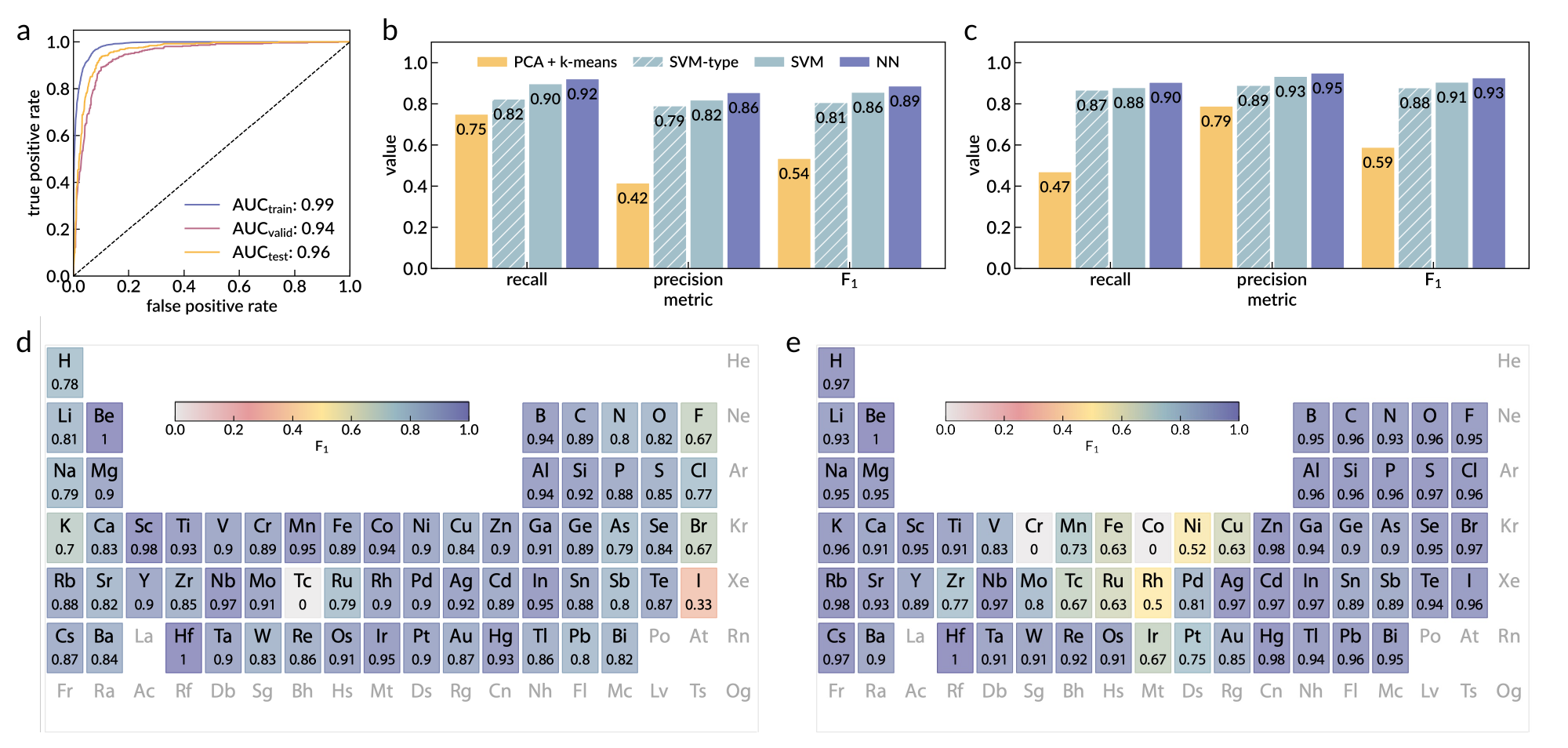}
  \caption{\raggedright\textbf{NN classifier performance.} (a) The receiver operating characteristic (ROC) curve showing the tradeoff between true and false positive rates for the NN model. The area under the curve (AUC) for each dataset is noted in the legend. (b-c) Comparative plots of the overall recall, precision, and F$_1$ scores for (b) topological and (c) trivial examples obtained using different methods discussed in the main text. (d-e) Element-specific F$_1$ scores for (d) topological and (e) trivial examples. Each element's entry lists its atomic number, atomic symbol, and F$_1$ score. Elements with no score listed were not present in the dataset.}
  \label{fig:figure3}
\end{figure*}

\subsection{Network architecture optimization}
The NN classifier inputs consist of the set of XANES spectra and atom types of each absorbing atom in a given material, as shown in \textbf{Figure~\ref{fig:figure2}a}, where atom types are encoded as one-hot feature vectors with a one at the index equaling the atomic number, and zeros elsewhere. The core–electron binding energy increases substantially with increasing atomic number, ranging from 284 eV for the C K-edge to 115,606 eV for the U K-edge \cite{penner2003x}, and thus representing the XANES spectra of all absorbers on a continuous energy scale would be either poorly resolved or exceedingly high-dimensional (\textbf{Figure~\ref{fig:figure2}a}). Separating the spectral and atom type information at the input facilitates the construction of element-specific channels and allows us to retain the spectral energy resolution. In addition to enabling the synthesis of information from different absorbers, a neural network comprises more complex, non-linear operations than PCA and thereby has the capability to learn more expressive representations of the input data. The network architecture is illustrated in \textbf{Figure~\ref{fig:figure2}b}. Fully-connected layers first operate on each spectral and atom-type input to obtain intermediate representations, termed the spectral and atom-type embeddings, respectively. The embedded spectra are assigned to element-specific channels through a direct product with the corresponding atom-type embedding. These composite features are subsequently added for a given material and flattened to a single array, which is passed to another series of fully-connected layers and activations that output the predicted binary topological class. Due to moderate class imbalance, samples were weighted to add greater penalty to the misclassification of topological examples.

\subsection{Machine learning model performance}
\textbf{Figure~\ref{fig:figure3}} summarizes the performance of the trained NN classifier. The receiver operating characteristic (ROC) curve, which indicates the tradeoff between true and false positive rates, is shown in \textbf{Figure~\ref{fig:figure3}a}. We use three different metrics in assessing the quality of prediction: recall, precision, and F$_1$ score. These metrics are defined as
\begin{subequations}
\begin{align}
    \text{recall:\enspace} r &= \frac{t_p}{t_p + f_n}, \\
    \text{precision:\enspace} p &= \frac{t_p}{t_p + f_p}, \\
    \text{F$_1$ score:\enspace} F_1 &= 2\frac{p\cdot{r}}{p + r},
\end{align}
\end{subequations}
where $t_p$ and $t_n$ denote the number of true positive and true negative predictions, and $f_p$ and $f_n$ denote the number of false positive and false negative predictions of a given class, respectively. The NN classifier achieved F$_1$ scores of $89\%$ and $93\%$ for topological and trivial classes, respectively. We compare these results to the performance of a traditional support vector machine (SVM) operating on one-hot encoded atom types only (denoted SVM-type) and on a concatenated array of spectra for all atom types (denoted SVM), as shown in \textbf{Figure~\ref{fig:figure3}b} and \textbf{c}. The average performance of the PCA and $k$-means clustering approach across all elements is also included for reference. Note that the concatenated feature vector input to the SVM contains zeros in place of spectra corresponding to elements not contained in the compound. We find that both the NN and SVM classifiers based on XANES spectral inputs outperform the baseline model relying on atom types alone, suggesting that XANES spectral features provide meaningful insight to topological indication. To maintain the same number of neurons between SVM-type and SVM models, the SVM-type inputs were copied 200 times (the length of the spectral inputs) to construct the input features, which led to a combined increase of $5\%$ in the $F_1$ scores compared to a minimal SVM-type model reported in \textbf{Figure~\ref{fig:figureS5}a} for comparison. The NN further improves upon the SVM model predictions, particularly in the precision of topological classification which increased by $4\%$. We note that the NN with both spectral and atom-type inputs achieves a combined improvement of $\sim7\%$ in the $F_1$ scores compared to a NN model of similar size operating on atom-type inputs alone (\textbf{Figure~\ref{fig:figureS5}a}). Additional details about the reference models are provided in the Supplemental Material. \hlc[cyan]{We also assess the sensitivity to the spectral energy resolution in} \textbf{Figure~\ref{fig:figureS7}}. \hlc[cyan]{While the main results of this work are obtained for spectra sampled at intervals of $\sim 0.28$ eV, we see that a sampling of $\sim 5$ eV is sufficient for comparable performance.} Finally, we compute the average metric scores obtained by the NN classifier individually for each absorbing element, shown in \textbf{Figure~\ref{fig:figure3}d} and \textbf{e} for topological and trivial examples, respectively. Corresponding results for the SVM model and additional plots for the NN classifier are shown in \textbf{Figure~\ref{fig:figureS4}} and \textbf{Figure~\ref{fig:figureS6}}, respectively. 

\begin{figure*}[!t]
  \centering
  \includegraphics[width=18cm]{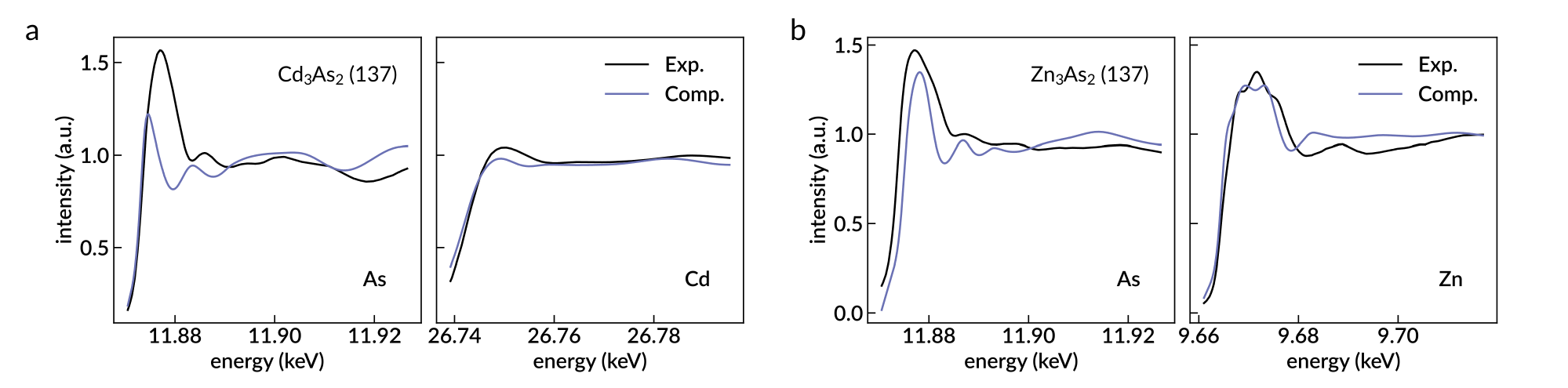}
  \caption{\raggedright\textbf{Comparison between experimental and computational XAS spectra.} \hlc[cyan]{Experimental (black) and computational (blue) K-edge XANES spectra of (a) As and Cd in Cd$_3$As$_2$ (topological) and (b) As and Zn in Zn$_3$As$_2$ (trivial). The spacegroup of each structure is indicated in parentheses. Both experimental and computational inputs in (a) and (b) are correctly classified.}}
  \label{fig:figure4}
\end{figure*}

\subsection{Application to experimental spectra}
\hlc[cyan]{While we are unable to include experimental spectra in our training set due to limited availability, we present a preliminary effort by making predictions on a small set of seven experimental XAS spectra and their computational counterparts, where available. The XAS experiments were performed at the 4-ID-D beamline of the Advanced Photon Source (APS) and include measurements of both topological and trivial compounds listed in Table}~\ref{tab:exp}. \hlc[cyan]{The predictions obtained using the experimental and computational spectra were all consistent with one another, though in one instance (MoSe$_2$) both are incorrectly classified, as shown in Table}~\ref{tab:exp}.\hlc[cyan]{Specifically, within this set of examples, the neural network correctly classifies six of the seven sets of experimental spectra, and five of the six sets of computational spectra (computational spectra were unavailable for one of the seven compounds). Additionally, we note that for the two ternary compounds, LaAlGe and CdGeYb, one of the three absorption edges could not be measured at this time; in these cases, the two available experimental spectra were used to make a prediction. As an example,} \textbf{Figure~\ref{fig:figure4}} \hlc[cyan]{shows the experimental and computational XAS spectra of the topological semimetal Cd$_3$As$_2$} (\textbf{Figure~\ref{fig:figure4}a}) \hlc[cyan]{and the isostructural trivial compound Zn$_3$As$_2$} (\textbf{Figure~\ref{fig:figure4}b}). \hlc[cyan]{While there is some misalignment of the experimental spectra relative to the computed ones, many of the key qualitative features are preserved. We expect that a certain tolerance in the misalignment is admissible, further reinforced by the results of the sensitivity analysis discussed in the previous section. Spectra for the remaining experimental examples are provided in the Supplementary Information as} \textbf{Figure~\ref{fig:figureS9}}.

\begin{table}[bh]
    \footnotesize
    \renewcommand{\arraystretch}{1.3}
    \centering
    \caption{\hlc[cyan]{Predictions on corresponding experimental and computational spectra}}
    \begin{threeparttable}
    \begin{tabular}{@{}p{0.5\textwidth}@{}}
    \centering
    \begin{tabular}{ccccc}
    	\hline\hline
    	& & \multicolumn{3}{c}{Class}\\
    	Material & Spacegroup & True & Pred. (Exp.) & Pred. (Comp.) \\
    	\hline
        NbAs & 109 & Topo. & Topo. & Topo. \\
        LaAlGe\tnote{a)} & 109 & Topo. & Topo. & Topo. \\
        Zn3As2 & 137 & Trivial & Trivial & Trivial \\
        Cd3As2 & 137 & Topo. & Topo. & Topo. \\
        CdGeYb\tnote{b)} & 189 & Topo. & Topo. & $\sim$ \\
        MoSe2 & 194 & Trivial & Topo.\tnote{d)} & Topo.\tnote{d)} \\
        CdTe & $\ast$\tnote{c)} & Trivial & Trivial & Trivial \\
    \bottomrule\addlinespace[1ex]
    \end{tabular}
    \end{tabular}
    \begin{tablenotes}
    \item[a)] Al K-edge was not measured.
    \item[b)] Yb K-edge was not measured.
    \item[c)] The same classifications are obtained for all computed spacegroups: 63, 152, 186, 216, and 225.
    \item[d)] Incorrectly predicted.
    \end{tablenotes}
    \end{threeparttable}
    \label{tab:exp}
\end{table}

\begin{table}[th]
    \footnotesize
    \renewcommand{\arraystretch}{1.3}
    \centering
    \caption{Predictions on mislabeled Weyl semimetals}
    \begin{threeparttable}
    \begin{tabular}{@{}p{0.5\textwidth}@{}}
    \centering
    \begin{tabular}{ccc}
    	\hline\hline
    	Material & Spacegroup & Predicted class \\
    	\hline
        TaAs & 109 & Topological \\
    	NbAs & 109 & Topological \\
        NbP & 109 & Topological \\
        WTe2 & 31 & Topological \\
        Ag2Se & 17 & Trivial \\
        LaAlGe & 109 & Topological \\
        Ba7Al4Ge9 & 42 & Topological \\
    	Cu2SnTe3 & 44 & Topological \\
        BiTeI & 143 & Trivial \\
        Al4Mo & 8 & Topological \\
        KOs2O6 & 216 & Topological \\
        Zn2In2S5 & 186 & Trivial \\
    \bottomrule\addlinespace[1ex]
    \end{tabular}
    \end{tabular}
    \end{threeparttable}
    \label{tab:weyl}
\end{table}

\section{Discussion}
Our results indicate that the NN classifier enables higher and more balanced predictive accuracy over the PCA and $k$-means clustering approach for a majority of elements, including significant improvement for alkali metals. Certain elements are better indicators of one class over another; for instance, the alkali metals and halogens appear to serve as somewhat poor indicators of topological samples but are well-predicted in trivial compounds. A possible explanation for this is that the elements in these columns rarely contribute to frontier orbitals (valence and conduction bands) in materials, and are thereby poor indicators of topology. Certain transition-metal elements, such as Cr, Co, Ni, Tc, and Rh, also exhibit imbalanced accuracy in the prediction of trivial and topological classes. This is most likely due to the overrepresentation of topological examples containing Cr, Co, Ni, and Rh (\textbf{Figure~\ref{fig:figureS5}c}), since accurate prediction of topological compounds is prioritized during training. Tc is the least abundant element in the dataset (\textbf{Figure~\ref{fig:figureS1}a} and \textbf{b}), which accounts for the model's weak performance on Tc-containing compounds. However, further investigation of the relevant spectroscopic features -- whether pre-edge, edge, or post-edge -- in connection with the corresponding electronic transitions (e.g. $1s \rightarrow 3d$) may be useful to better understand performance barriers for transition metals. Finally, we comment on the comparatively low precision obtained for topological over trivial examples, $86\%$ and $95\%$, respectively. While the higher false positive rate of topological materials may suggest additional model improvements are needed, it may also indicate missed topological candidates. In fact, since the TQC formalism considers only the characters of electronic bands at high-symmetry points, it may incorrectly classify certain Weyl semimetals with topological singularities at arbitrary $k$-points ~\cite{vergniory2019complete}. In particular, we identified 12 experimentally-verified \cite{yan2017topological} or theoretically-predicted Weyl semimetals \cite{xu2020comprehensive} that are labeled as trivial in the TQC database, 9 of which we correctly predict as topological using our NN classifier (Table \ref{tab:weyl}). Thus, the potential presence of topological singularities not considered in the TQC formalism might account for some loss of precision in the classification of topological examples. In addition, we summarize in Supplementary Table~\ref{tab:TableS1} the top 100 predicted topological materials from a collection of 459 samples not represented in the TQC database. These are the top candidates predicted by our model that may contain topological singularities. We do note that the success of the NN classifier can be attributed significantly to the presence of particular elements; further work is being pursued to more accurately decouple this contribution from that of more subtle variations in the XAS spectral features for a given absorbing element.

\section{Conclusion}
We explored the predictive power of XAS as a potential discriminant of topological character by training and evaluating a NN classifier on more than $10,000$ examples of computed XANES spectra~\cite{mathew2018high} labeled according to the largest catalogue of topological materials~\cite{vergniory2019complete,vergniory2021all}. A number of important extensions are envisioned for this work, such as its application to experimental XANES data, incorporation of a multi-fidelity approach to favor experimentally validated examples~\cite{meng2020composite}, expansion of the energy range to the extended X-ray absorption fine structure (EXAFS) regime, and inquiry into the detailed contribution from spectral features for individual elements. The theoretical connection between band topology and the local chemical environment encoded in XANES spectra has not yet been established, and we envision data-driven methods as a possible tool in aiding this theoretical development. Our current results demonstrate a promising pathway to develop robust experimental protocols for high-throughput screening of candidate topological materials aided by machine learning methods. Additionally, the flexibility of the XAS sample environment can further enable the study of materials whose topological phases emerge when driven by electric, magnetic, or strain fields, and even presents the opportunity to study topology with strong disorder and topology in amorphous materials~\cite{PhysRevLett.118.236402,prodan2011disordered}. Thus, machine learning-empowered XAS may be poised to become a simple but powerful experimental tool for topological classification.

\section{Methods}
\noindent \textbf{Data processing}
The computed XANES spectra of each absorbing atom were interpolated and re-sampled at $200$ evenly-spaced energy values. Each XANES spectrum spanned an energy range of $56$ eV, and spectra from the same absorbing atom were co-aligned using the calculated absolute energy scale. Spectra of the same absorbing atom were standardized by centering the mean of the average intensities over the sampled energy range, and scaling by the mean of the standard deviations in intensity values. \\

\noindent \textbf{Machine learning}
Principal component analysis and SVM model implementation and training were carried out using the scikit-learn Python library \cite{scikit-learn}. The NN models presented in this work were implemented in Python using the PyTorch \cite{NEURIPS2019_9015} and PyTorch Geometric \cite{Fey/Lenssen/2019} libraries. The atom-type embeddings were obtained using a single fully-connected layer with 93 input and output neurons. The spectral embeddings of the original 200-feature spectra were obtained using a series of two fully-connected layers with 256 and 64 output neurons, respectively, each followed by a dropout layer with a rate of 0.5 and a rectified linear unit (ReLU) activation. The composite embedded features had dimensions of 5952 and were passed to a second series of two fully-connected layers with 256 and 64 output neurons, respectively, each followed by a dropout layer with a rate of 0.5 and a ReLU activation. A final, sigmoid-activated, fully-connected layer was then used to output the scalar prediction. The models were trained on a Quadro RTX 6000 graphics processing unit (GPU) with $24\,\text{GB}$ of random access memory (RAM). Optimization was performed using the Adam optimizer to minimize the binary cross-entropy loss. \\

\noindent \textbf{Sample preparation}.
\hlc[cyan]{NbAs and CdTe crystals were grown using chemical vapor transport while LaAlGe, CdGeYb, Cd$_3$As$_2$, Zn$_3$As$_2$ and MoSe$_2$ crystals were grown using the flux method as described in literature. The samples exhibited clear, lustrous surfaces with demarcated straight edges indicating the orientation of the crystal axes. The samples were not polished.} \\

\noindent \textbf{X-ray absorption spectroscopy}
\hlc[cyan]{XAS experiments were performed at the 4-ID-D beamline of the Advanced Photon Source, Argonne National Laboratory. The X-ray energy was selected using a Si (111) double crystal monochromator, which was detuned to reject harmonics. Measurements were recorded near the K absorption edge for each element. For absorption edges below 23 keV, a Pd mirror was employed to further reject harmonics. Measurements were done at room temperature and in transmission mode using N$_2$ and Ar filled ion chambers to detect both incident and transmitted intensities, respectively. Prior to making predictions, experimental spectra were pre-processed as follows. First, a linear background was fit to the pre-edge region and subtracted. The resulting spectra were fit with an arctangent function of the form $a_1\left(1 + 2 \tan^{-1}\left(a_2(E - a_3)\right)/\pi\right)/2$ with fitting parameters $\{a_i\}$ and measured energies $E$, and subsequently scaled by $1/a_1$. This ensured that experimental intensities were scaled consistently with computational ones, which are 0 at energies well below the absorption edge and approach 1 at energies well above the absorption edge. Finally, the experimental spectra were shifted in energy so that the area under the fitted arctangent matched that of the arctangent fit to the average computational spectrum for each absorbing atom. Examples of these pre-processing steps are shown in} \textbf{Figure~\ref{fig:figureS10}} \hlc[cyan]{of the Supplemental Material. Finally, experimental spectra were interpolated and scaled according to the means and standard deviations of the computational spectra as described in the Data processing section.} \\

\begin{acknowledgments}
N.A.~acknowledges National Science Foundation GRFP support under Grant No.~1122374.
J.A.~acknowledges National Science Foundation GRFP support under Grant No.~DGE-1745303.
N.A. and M.L. acknowledge the support from the U.S. Department of Energy (DOE), Office of Science (SC), Basic Energy Sciences (BES), Award No. DE-SC0021940. F.H., T.N. and M.L. acknowledge the support from the DOE Award No. DE-SC0020148. M.L. is partially supported by NSF DMR-2118448, the Norman C. Rasmussen Career Development Chair, and the Class of 1947 Career Development Chair, and acknowledges the support from Dr. R. Wachnik.
B.A.B. and N.R. gratefully acknowledge financial support from the Schmidt DataX Fund at Princeton University made possible through a major gift from the Schmidt Futures Foundation, NSF-MRSEC Grant No. DMR-2011750 and the European Research Council (ERC) under the European Union’s Horizon 2020 research and innovation program (Grant Agreement No. 101020833).
C.H.R.~was partially supported by the Applied Mathematics Program of the
U.S. DOE Office of Science Advanced Scientific Computing Research under
Contract No.~DE-AC02-05CH11231.
Work performed at the Center for Nanoscale Materials, a U.S. Department of Energy Office of Science User Facility, was supported by the U.S. DOE, Office of Basic Energy Sciences, under Contract No. DE-AC02-06CH11357. This material is based, in part, upon work supported by Laboratory Directed Research and Development (LDRD) funding from Argonne National Laboratory, provided by the Director, Office of Science, of the U.S. Department of Energy under Contract No. DE-AC02-06CH11357. \hlc[cyan]{This research used resources of the Advanced Photon Source, a U.S. Department of Energy (DOE) Office of Science User Facility operated for the DOE Office of Science by Argonne National Laboratory under Contract No. DE-AC02-06CH11357.}\\
\end{acknowledgments}

\noindent \textbf{Competing interests} The authors declare no competing interests. \\

\noindent \textbf{Additional information}
Supplemental Material is available for this paper.
Correspondence and requests for materials should be addressed
to Nina Andrejevic (nandrejevic@anl.gov), Mingda Li (mingda@mit.edu), or Chris H.\ Rycroft (chr@math.wisc.edu). \\

\noindent \textbf{Data availability}
All the data and code supporting the findings are available from the corresponding authors upon reasonable request.

\bibliography{main}

\clearpage
\clearpage
\setcounter{section}{0}
\setcounter{equation}{0}
\setcounter{figure}{0}
\setcounter{table}{0}
\setcounter{page}{1}
\makeatletter
\renewcommand{\theequation}{S\arabic{equation}}
\renewcommand{\thefigure}{S\arabic{figure}}
\renewcommand{\thetable}{S\arabic{table}}

\onecolumngrid
\begin{center}
    \large\textbf{Supplemental material for ``Machine learning spectral indicators of topology"} \\
    \vspace{\baselineskip}
    \normalsize Nina Andrejevic,\textit{$^{1,2,3,*}$} Jovana Andrejevic,\textit{$^{4,5,*}$} B. Andrei Bernevig,\textit{$^{6,7,8}$}
    Nicolas Regnault,\textit{$^6$} Fei Han,\textit{$^{2,9}$} \\ Gilberto Fabbris,\textit{$^{10}$} Thanh Nguyen,\textit{$^{2,9}$} Nathan C. Drucker,\textit{$^{2,5}$} Chris H.~Rycroft,\textit{$^{11,5,12,\dag}$} and Mingda Li \textit{$^{2,9,\ddag}$} \\
    \textsuperscript{1}\textit{Center for Nanoscale Materials, Argonne National Laboratory, Lemont, IL 60439, USA} \\
    \textsuperscript{2}\textit{Quantum Measurement Group, Massachusetts Institute of Technology, Cambridge, MA 02139, USA} \\
    \textsuperscript{3}\textit{Department of Materials Science and Engineering,} \\
    \textit{Massachusetts Institute of Technology, Cambridge, MA 02139, USA} \\
    \textsuperscript{4}\textit{Department of Physics, University of Pennsylvania, Philadelphia, PA 19104, USA} \\
    \textsuperscript{5}\textit{John A.~Paulson School of Engineering and Applied Sciences,} \\
    \textit{Harvard University, Cambridge, MA 02138, USA} \\
    \textsuperscript{6}\textit{Department of Physics, Princeton University, Princeton, NJ 08544, USA} \\
    \textsuperscript{7}\textit{Donostia International Physics Center, P. Manuel de Lardizabal 4, 20018 Donostia-San Sebastian, Spain} \\
    \textsuperscript{8}\textit{IKERBASQUE, Basque Foundation for Science, Bilbao, Spain} \\
    \textsuperscript{9}\textit{Department of Nuclear Science and Engineering,} \\
    \textit{Massachusetts Institute of Technology, Cambridge, MA 02139, USA} \\
    \textsuperscript{10}\textit{Advanced Photon Source, Argonne National Laboratory, Lemont, IL 60439, USA} \\
    \textsuperscript{11}\textit{Department of Mathematics, University of Wisconsin-Madison, Madison, WI 53706, USA} \\
    \textsuperscript{12}\textit{Computational Research Division, Lawrence Berkeley Laboratory, Berkeley, CA 94720, USA} \\
    
\end{center}

\begin{figure*}[!ht]
  \centering
  \includegraphics[width=16cm]{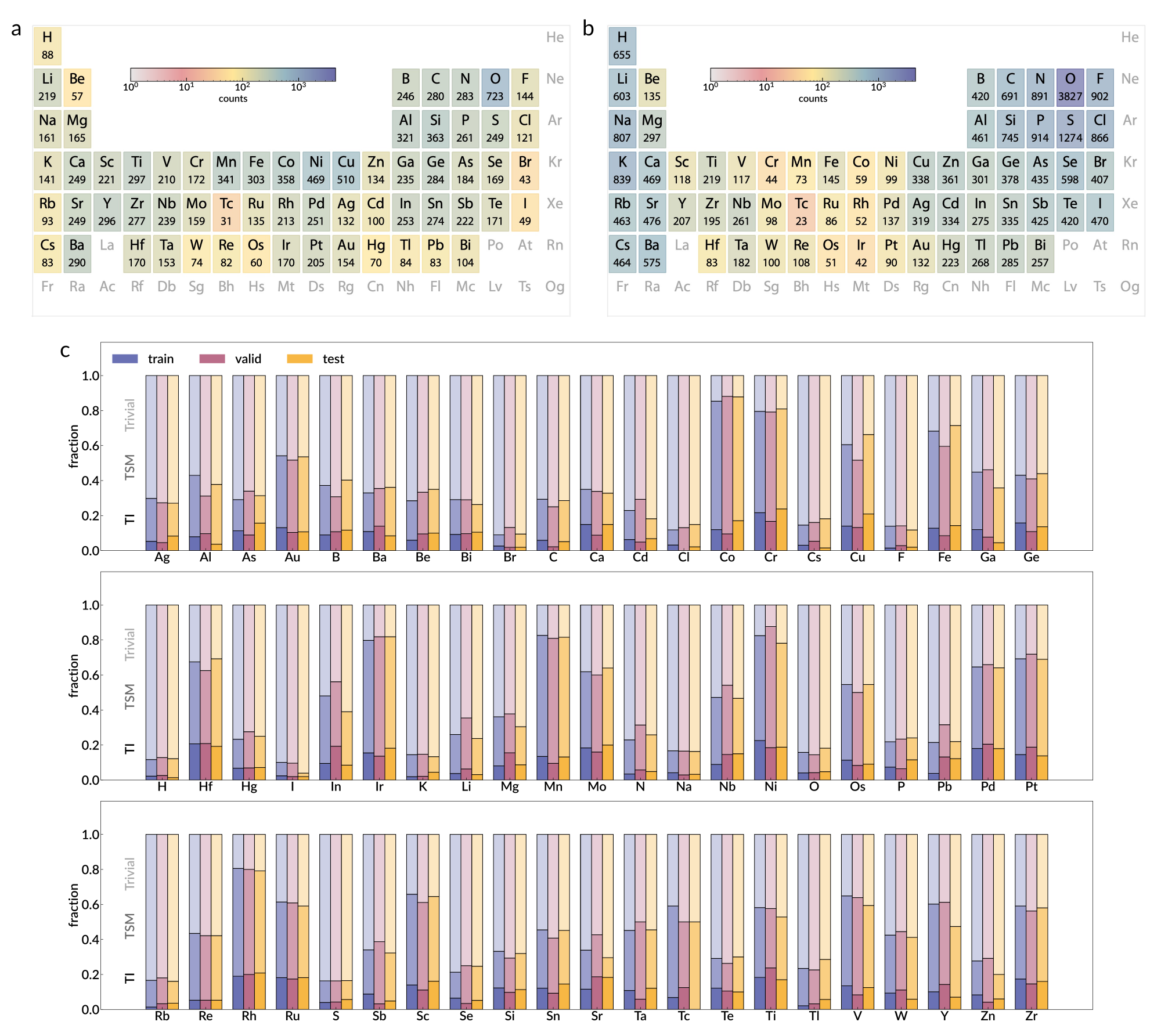}
  \caption{\raggedright\textbf{Element and topological class frequencies in the dataset.} (a) The total number of topological samples across training, validation, and testing data containing each element. Each element's entry includes its atomic number, atomic symbol, and number of samples, and is colored by the number of samples. (b) The total number of trivial samples by element. Elements with no counts listed were not present in the dataset. (c) The fraction of topological and trivial samples, by element, in the training, validation, and testing sets. The data subdivision reflects a balanced representation of absorbing elements and topological class across the datasets. TI and TSM denote topological insulators and topological semimetals, respectively.}
  \label{fig:figureS1}
\end{figure*}

\clearpage
\begin{sidewaysfigure}[!ht]
  \centering
  \includegraphics[width=22cm]{figureS2.png}
  \caption{\raggedright\textbf{Complete results of PCA and \textit{k}-means clustering.} Decision boundary visualizations of classifications by unsupervised \textit{k}-means clustering for all elements. The clusters are visualized along the first ($x$-axis) and second ($y$-axis) principal components in the scatter plots. Scattered points are colored according to their true class: topological (orange) or trivial (blue). The background is shaded according to the cluster-assigned class.}
  \label{fig:figureS2}
\end{sidewaysfigure}
\clearpage

\clearpage
\begin{sidewaysfigure}[!ht]
  \centering
  \includegraphics[width=22cm]{figureS3.png}
  \caption{\raggedright\textbf{Average XANES K-edge spectra by element and topological class.} Visualization of the average XANES spectra for each absorbing atom present in the dataset used for this work, separated by topological (orange) and trivial (blue) class. Shaded regions correspond to one standard deviation.}
  \label{fig:figureS3}
\end{sidewaysfigure}
\clearpage

\begin{figure*}[!ht]
  \centering
  \includegraphics[width=16cm]{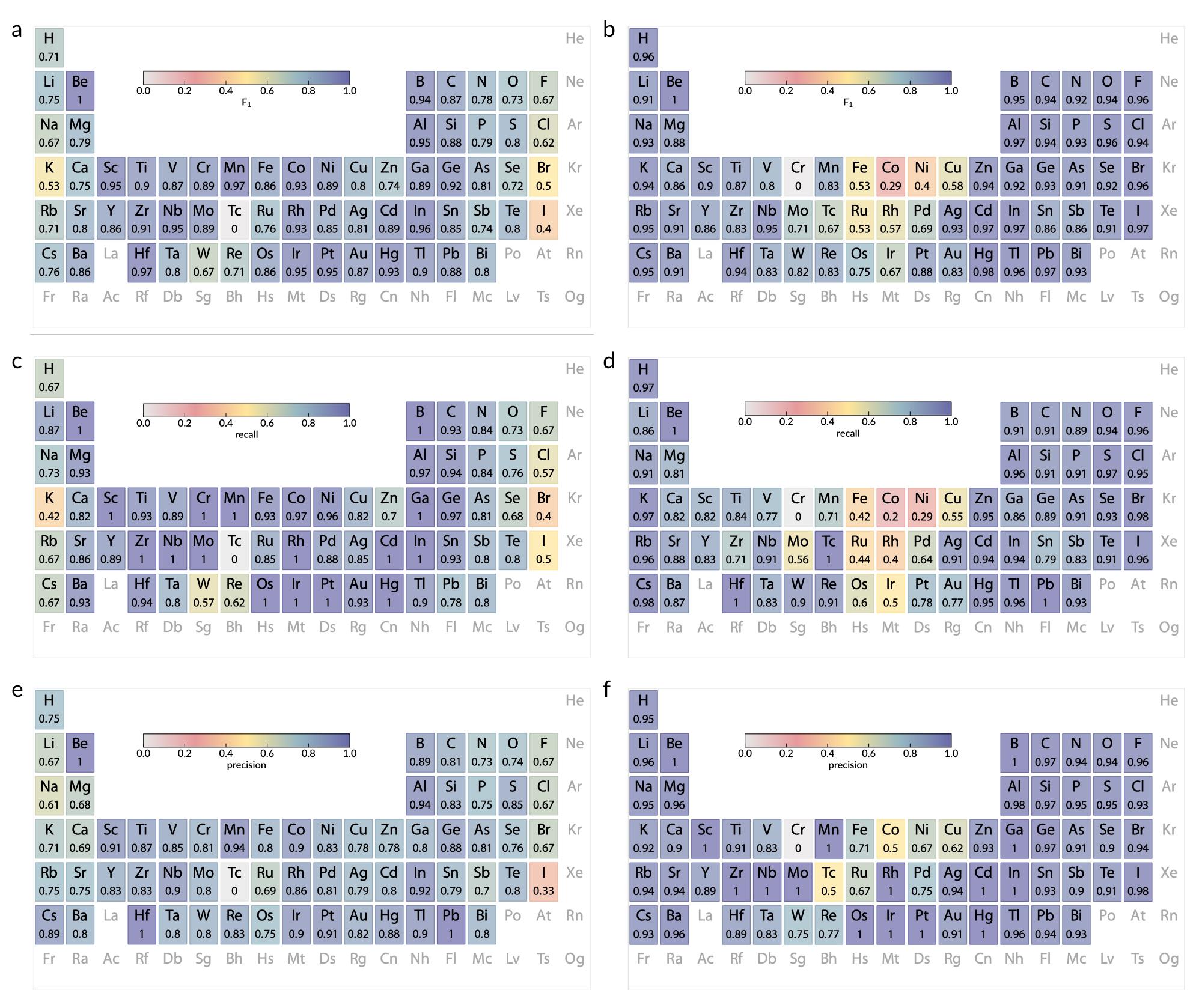}
  \caption{\raggedright\textbf{Support vector machine (SVM) model performance.} Element specific (a-b) F$_1$ scores, (c-d) recall, and (e-f) precision for topological (left column) and trivial (right column) examples, respectively. Elements with no score listed were not present in the dataset.}
  \label{fig:figureS4}
\end{figure*}

\begin{figure*}[!ht]
  \centering
  \includegraphics[width=18cm]{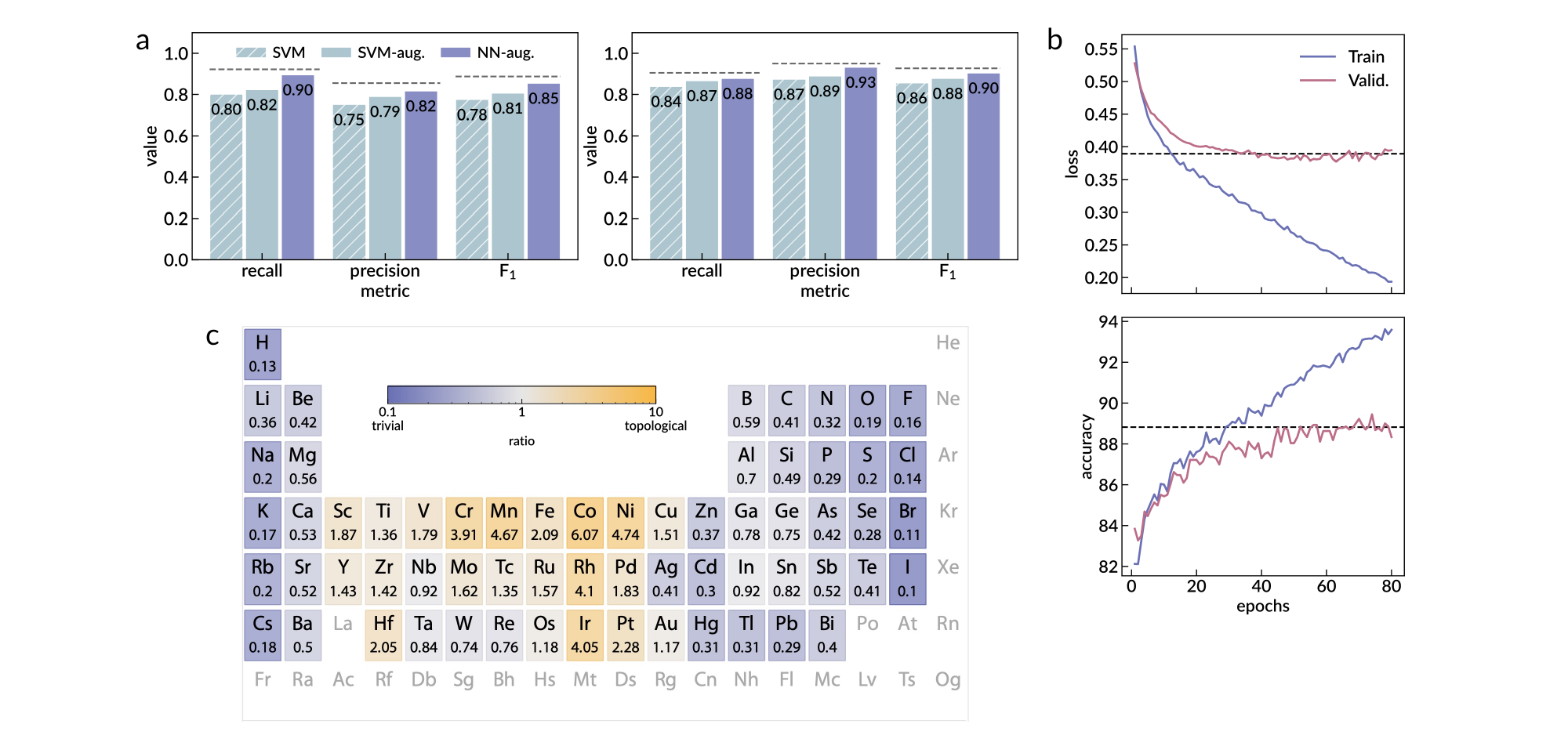}
  \caption{\raggedright\textbf{Additional model performance metrics.} (a) Comparative plots of the overall recall, precision, and F$_1$ scores for (left) topological and (right) trivial examples obtained using only atom-type inputs. As noted in the main text, augmented inputs consisting of 200 copies of the one-hot encoded atom types were passed to the SVM to maintain the same number of neurons between SVM-type and SVM models. In the barplots shown, SVM and SVM-aug. refer to SVM models operating on the original and augmented atom-type inputs, respectively. NN-aug. refers to the neural network model without the spectral embedding layers; instead, the direct product is performed between the atom-type embedding and an array of ones equal in length to the spectral embedding vector. Dashed gray lines indicate the scores of the full NN model reported in the main text. (b) Representative training history of the full NN model, indicating the loss and accuracy at the early stopping point determined using the validation set (black dashed lines). (c) Ratios of topological to trivial examples present in the dataset for each absorbing atom, highlighting the overrepresentation of topological examples containing certain transition metals.}
  \label{fig:figureS5}
\end{figure*}

\begin{figure*}[!ht]
  \centering
  \includegraphics[width=16cm]{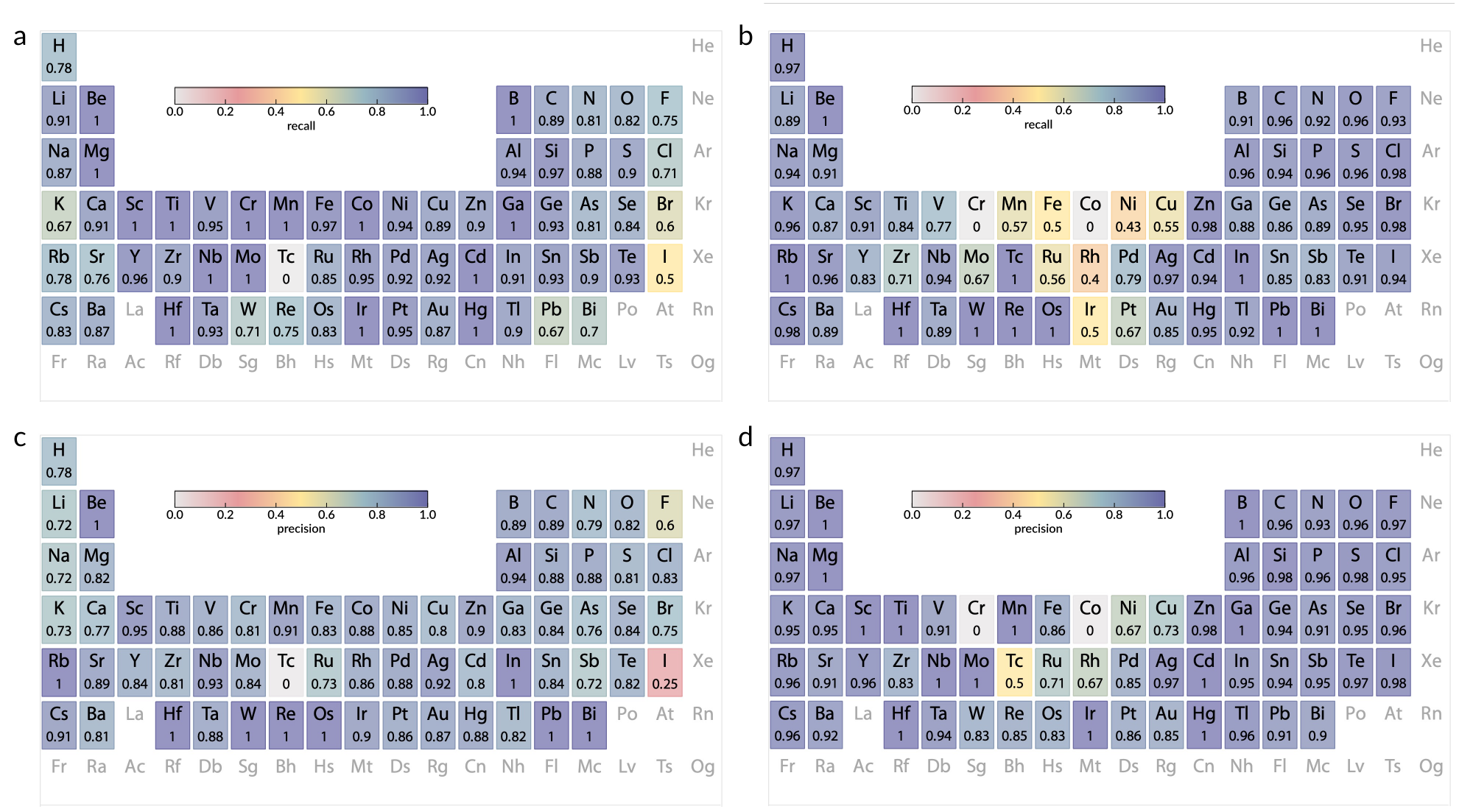}
  \caption{\raggedright\textbf{Neural network (NN) classifier recall and precision.} Element specific (a-b) recall and (c-d) precision for topological (left column) and trivial (right column) examples, respectively. Elements with no score listed were not present in the dataset.}
  \label{fig:figureS6}
\end{figure*}

\begin{figure*}[!ht]
  \centering
  \includegraphics[width=18cm]{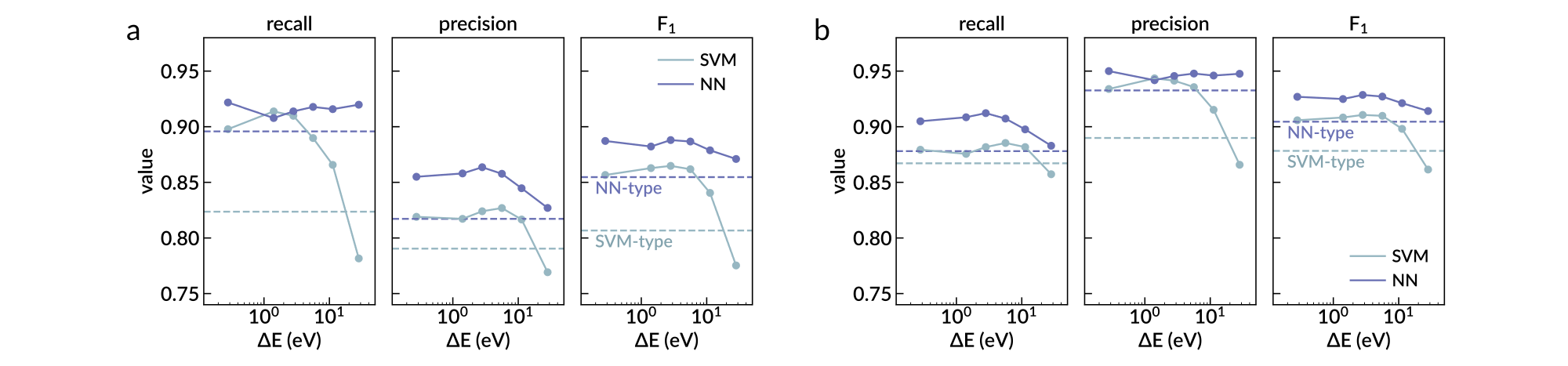}
  \caption{\raggedright\textbf{Sensitivity to spectral energy resolution.} \hlc[cyan]{The overall recall, precision, and F$_1$ scores for (a) topological and (b) trivial examples as a function of the energy interval $\Delta{E}$ between sampled points of the XANES spectra. Scores are presented for both the SVM and NN models, with scores from the atom-type only models (SVM-type and NN-type) shown as a reference by the dotted lines. Spectra were resampled at lower resolutions by computing their average values over length $\Delta{E}$ intervals along the energy axis for varied $\Delta{E}$. To maintain the same number of neurons across all resolutions, the averaged values were copied by the number of original samples within each interval such that all spectral inputs have length 200.}}
  \label{fig:figureS7}
\end{figure*}

\begin{figure*}[!ht]
  \centering
  \includegraphics[width=18cm]{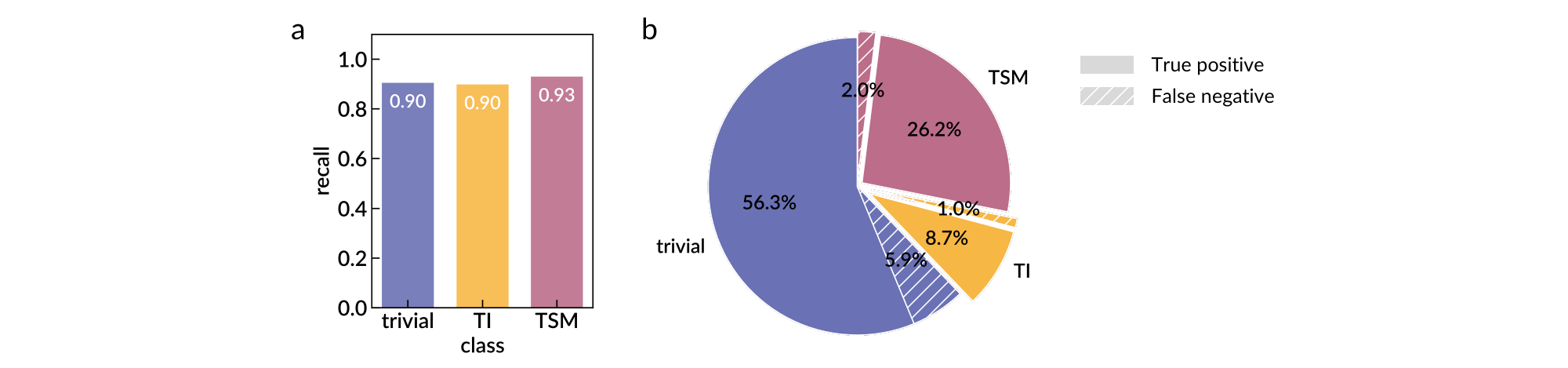}
  \caption{\raggedright\textbf{Neural network classifier performance per subclass.} \hlc[cyan]{(a) Recall scores for each subclass -- trivial, topological insulator (TI), and topological semimetal (TSM) -- showing comparable performance. (b) Proportion of correctly (solid) and incorrectly (hatched) classified examples within each subclass: Trivial (blue), TI (yellow), and TSM (pink).}}
  \label{fig:figureS8}
\end{figure*}

\begin{figure*}[!ht]
  \centering
  \includegraphics[width=18cm]{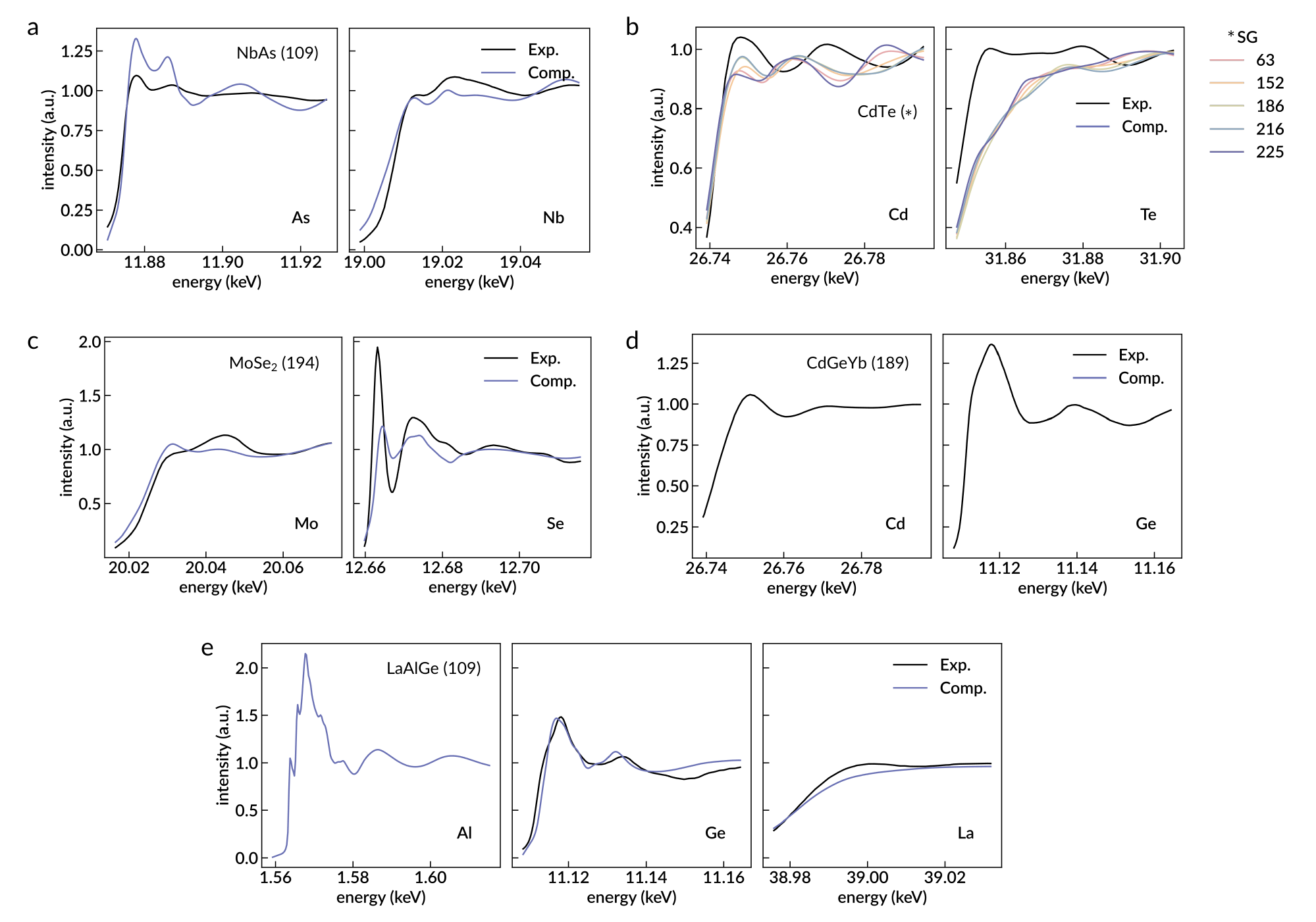}
  \caption{\raggedright\textbf{Additional experimental and corresponding computational XAS spectra.} \hlc[cyan]{Experimental (black) and computational (color) K-edge XANES spectra of select examples: (a) As and Nb in NbAs (topological), (b) Cd and Te in CdTe (trivial), (c) Mo and Se in MoSe$_2$ (trivial), (d) Cd and Ge in CdGeYb (topological), and (e) Al, Ge, and La in LaAlGe (topological). The spacegroup of each structure is indicated in parentheses; for (b), five computational spectra corresponding to different spacegroups but same topological class were present. We note that corresponding computational spectra were not available in (d), and the experimental spectra for Yb in (d) and Al in (e) were not measured. NN classifier predictions on the experimental and computational spectra are provided in Table}~\ref{tab:exp}\hlc[cyan]{ of the main text.}}
  \label{fig:figureS9}
\end{figure*}

\begin{figure*}[!ht]
  \centering
  \includegraphics[width=18cm]{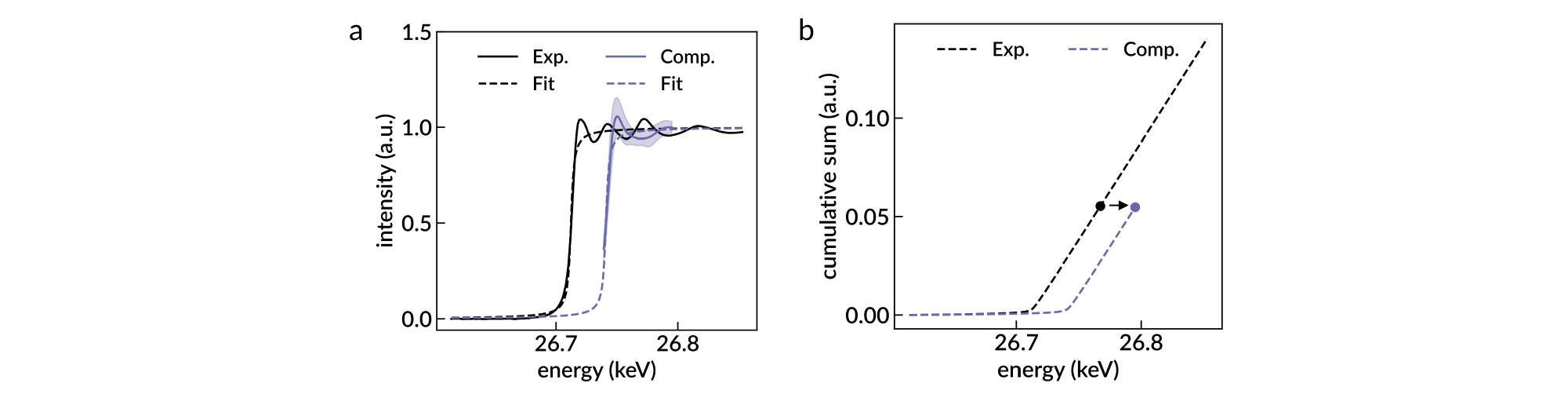}
  \caption{\raggedright\textbf{Scaling and alignment of experimental spectra.} \hlc[cyan]{(a) Representative experimental XAS spectrum (black solid line) of Cd after background subtraction and intensity scaling. Scaling is performed by fitting each experimental spectrum with an arctangent of the form $a_1\left(1 + 2 \tan^{-1}\left(a_2(E - a_3)\right)/\pi\right)/2$ with fitting parameters $\{a_i\}$ and measured energies $E$, and subsequently scaled by $1/a_1$. The re-scaled arctangent (reduced by the factor of $1/a_1$) fitted to the experimental spectrum is also shown (black dashed line). The resulting intensities are consistent with those of the computational spectra, as seen by comparison to the average computational spectrum for Cd (blue solid line). Light blue shading represents one standard deviation away from the mean intensity. The arctangent fit to the average computational spectrum is also shown (blue dashed line). The absorption edges are typically not aligned due to slight differences between the measured energies and the absolute energy scale adopted in the high-throughput calculations. (b) The cumulative sum of the arctangents shown in (a) for the experimental (black) and computational (blue) data. The experimental spectrum is shifted in energy as indicated by the black arrow, at which point the total integrated areas of the two arctangents match.}}
  \label{fig:figureS10}
\end{figure*}

\clearpage
\begin{table}
\caption{\raggedright\textbf{Top predicted topological materials.} List of 100 unclassified examples with the highest predicted score for topological class, all exceeding 0.90. Indicated are the full chemical formula, spacegroup (SG), ICSD identifier, and Materials Project identifier (mp-ID).}
\label{tab:TableS1}
\begin{subtable}[t]{.5\textwidth}
\begin{tabular}{@{\extracolsep{10pt}}llll}
\toprule
           Formula &  SG &   ICSD &      mp-ID \\
\midrule
        Cr8 B8 Ir4 &          63 & 418162 &  mp-569911 \\
          Ta4 Pt12 &          11 & 105804 &  mp-567638 \\
   Sc4 Mn2 B4 Rh10 &         127 &  51436 &  mp-542022 \\
        Sc3 Al1 C1 &         221 &  50161 &    mp-4079 \\
       Zr2 Fe12 P7 &         174 &  25757 &  mp-540809 \\
        Hf18 V8 S2 &         194 &  81822 &   mp-17105 \\
       In2 Ni21 B6 &         225 & 380216 &   mp-21551 \\
            V4 Sn8 &          70 & 106551 &   mp-20887 \\
       Ba6 Al6 Ga4 &         194 & 415931 &  mp-569218 \\
         Y3 In1 C1 &         221 &  80955 &   mp-19817 \\
           B8 Rh10 &         194 &  86395 &  mp-567926 \\
       Mn1 Ga1 Ni2 &         225 & 188992 &   mp-20228 \\
          Ga18 Rh4 &           7 & 414305 &   mp-31312 \\
           Ti1 Se2 &         164 &  80091 &    mp-2194 \\
               Sc2 &         194 & 164088 &      mp-67 \\
           Mg8 Cu4 &          70 & 174176 &    mp-2481 \\
           Mg2 Cu4 &         227 &  46007 &    mp-1038 \\
        Cu1 Sn1 F6 &           2 &  36514 &    mp-4701 \\
         Hg10 Au12 &         193 &  58475 &    mp-1812 \\
           Mn6 As4 &          12 &  75510 &   mp-28916 \\
          Mn12 As8 &          36 &  73251 &  mp-568856 \\
           Zr1 B12 &         225 & 409635 &    mp-1084 \\
           Y12 Co8 &          58 &   1844 &  mp-616512 \\
            Mo8 C4 &          60 & 246146 &    mp-1552 \\
       Mn1 Sb2 F12 &           2 & 411522 &  mp-555052 \\
        Zr4 Co2 P2 &          11 &  84825 &   mp-29152 \\
      Ba8 Mn4 Se12 &          62 &  26230 &   mp-18513 \\
      Sc6 Si7 Ni16 &         225 & 159262 &    mp-5382 \\
            Nb1 S2 &          42 &  72725 &    mp-2648 \\
           Ti6 Ni8 &         148 & 166371 &  mp-567653 \\
Ba2 Ca1 Tl2 Cu2 O8 &         123 &  68198 &  mp-573069 \\
           Ti2 Ni2 &          11 & 164155 &    mp-1048 \\
      Rb8 Co4 Br16 &          62 & 280362 &   mp-23366 \\
          Ba6 Pb10 &          63 & 165184 &  mp-630923 \\
       Sc8 N2 Cl12 &          55 & 201978 &   mp-23414 \\
  Mn1 Hg1 C4 S4 N4 &          82 &  87889 &  mp-541914 \\
       Sc4 Si4 Cu4 &          62 &  86391 &   mp-20083 \\
       Sc7 B1 Cl12 &         148 & 201975 &  mp-505489 \\
           Ba2 Cd4 &          74 & 260668 &   mp-11266 \\
       Ba1 Ni2 As2 &         139 & 164197 &  mp-568280 \\
          Ni16 B12 &          62 &  24307 &  mp-640067 \\
          Sc8 Te12 &          70 & 171145 &   mp-12383 \\
          Hf6 Ni14 &           2 &   2417 &   mp-27166 \\
       Mn2 Cr2 F10 &          15 &    939 &  mp-555156 \\
       Y4 Ge10 Ru6 &          72 &  86498 &  mp-620812 \\
       Mn2 Se4 O10 &          15 &  73936 &   mp-19175 \\
        Mo4 P4 Ru4 &          62 &  98405 &   mp-22451 \\
           Mg2 Pd6 &         139 & 150228 &   mp-12742 \\
          Ba6 Sn10 &          63 & 167166 &   mp-18636 \\
         Y1 B2 Ir3 &          12 &  99236 &   mp-31098 \\
\bottomrule
\end{tabular}       
\end{subtable}\hfill
\begin{subtable}[t]{.5\textwidth}       
\begin{tabular}{@{\extracolsep{10pt}}llll}
\toprule
           Formula &  SG &   ICSD &      mp-ID \\
\midrule
   Ba2 Cu2 Te4 O14 &          40 & 404297 &  mp-557735 \\
        Mg3 Ni9 B6 &         181 & 156018 &  mp-571428 \\
            V3 Se4 &          12 &  84195 &   mp-22700 \\
        Ti4 Cu2 S8 &         227 &  35269 &    mp-3951 \\
      Ba4 Fe8 Se12 &          62 &  16308 &  mp-504563 \\
          Cd20 Cu8 &         194 &  58958 &   mp-30696 \\
       Mg7 Ti1 H16 &         225 & 152886 &  mp-644281 \\
   Ba6 Na2 Ir4 O18 &         194 & 413452 &   mp-21876 \\
         K1 Cr5 S8 &          12 &   2568 &   mp-12178 \\
       Nb6 Sn2 S12 &         182 &  83038 &    mp-9407 \\
       K8 Co4 Cl16 &          14 &    661 &  mp-571314 \\
           Sc12 P8 &          62 &  41678 &   mp-22600 \\
           Co2 Te4 &          58 &  86115 &    mp-9945 \\
           Sc12 P8 &          62 &  41679 &   mp-21182 \\
        Ba1 Fe2 P2 &         139 &  10468 &    mp-4883 \\
    Mn2 H4 Se2 O10 &          15 &  66746 &  mp-643412 \\
        Rb2 Fe4 S6 &          63 &  99505 &    mp-3787 \\
             V2 P2 &         194 &  42444 &    mp-1114 \\
            Y8 C14 &          14 &  86049 &    mp-9530 \\
           Nb4 Sn8 &          70 & 106552 &    mp-1046 \\
            Ni3 S2 &         155 &  23114 &     mp-362 \\
   Ba6 Ca2 Ir4 O18 &          15 & 245253 &   mp-17448 \\
           Sr8 In8 &          43 & 414234 &  mp-655461 \\
      Mn4 Ga8 Te16 &          62 &  67402 &  mp-653120 \\
               Ba2 &         165 & 280680 & mp-1008283 \\
          Sn16 Pd8 &         142 & 413281 &    mp-1573 \\
       Ta6 Pb2 S12 &         182 &  83037 &   mp-20784 \\
      Ba6 Fe6 Se14 &         186 &  16310 &  mp-567429 \\
        K12 Co2 S8 &         186 &  68603 &   mp-14794 \\
       Ca4 Pb2 Au4 &         127 & 409531 &   mp-20723 \\
        Tl1 Cr5 S8 &          12 &  40821 &  mp-541823 \\
       Mn1 Ga2 Se4 &          82 &  85674 &   mp-20261 \\
         K4 Mn6 S8 &          13 & 411172 &   mp-29861 \\
          Sn12 Pd4 &          64 & 413279 &    mp-1371 \\
           Fe2 Te4 &          58 &  86518 &   mp-19880 \\
         K2 Mn2 F6 &          11 &  89343 &  mp-644332 \\
    Cs4 Co2 Si2 O8 &          36 &  93878 &  mp-542168 \\
      Zn10 B6 Rh14 &          51 & 107495 &  mp-510697 \\
           Ni3 Sn1 &         221 & 181127 &   mp-11522 \\
   Cu1 H8 C6 N6 O2 &           2 & 156682 &  mp-583429 \\
           Fe10 C4 &           2 & 181367 &  mp-645339 \\
      Mg6 Si16 Ir6 &         216 & 416868 &  mp-569313 \\
        Ba5 V5 O14 &         164 &  78163 &   mp-19088 \\
             V4 O6 &         167 &   1473 &   mp-18937 \\
       Y4 Ge10 Ir6 &          72 &  86503 &  mp-541609 \\
        Co2 C8 N12 &          58 &  85618 &   mp-22403 \\
       K12 Co2 Se8 &         186 &  68605 &   mp-14795 \\
           Ag8 F20 &           2 &  95832 &  mp-542298 \\
        K12 Mn2 S8 &         186 &  65448 &   mp-18244 \\
         Ga10 Fe12 &          12 &   2146 &  mp-570920 \\
\bottomrule
\end{tabular}
\end{subtable}
\end{table}

\end{document}